\documentclass[11pt]{article}
\usepackage{titlesec}
\titleformat{\paragraph}[runin]{\normalfont\itshape}{\theparagraph.}{.3em}{}[.]\titlespacing{\paragraph}{0pt}{1ex plus .1ex minus .2ex}{.5em}
\usepackage{amsmath,amssymb,bm}
\usepackage{mathabx}
\usepackage[T1]{fontenc}
\usepackage[utf8]{inputenc}
\usepackage{lmodern}
\usepackage{mathtools}
\usepackage{dsfont}
\pdfoutput=1
\usepackage[english]{babel}
\usepackage[letterpaper, hmargin=1in, top=1in, bottom=1.2in, footskip=0.6in]{geometry}
\usepackage{graphicx} 
\usepackage{booktabs} 
\usepackage{color}
\definecolor{aquamarine}{rgb}{0.5, 1.0, 0.83}
\definecolor{ao(english)}{rgb}{0.0, 0.5, 0.0}
\definecolor{armygreen}{rgb}{0.29, 0.33, 0.13}
\definecolor{awesome}{rgb}{1.0, 0.13, 0.32}
\definecolor{ballblue}{rgb}{0.13, 0.67, 0.8}
\definecolor{bittersweet}{rgb}{1.0, 0.44, 0.37}
\definecolor{blue}{rgb}{0.0, 0.0, 1.0}
\definecolor{brinkpink}{rgb}{0.98, 0.38, 0.5}
\definecolor{ballblue}{rgb}{0.13, 0.67, 0.8}
\definecolor{brightturquoise}{rgb}{0.03, 0.91, 0.87}
\definecolor{blue-green}{rgb}{0.0, 0.87, 0.87}
\definecolor{caribbeangreen}{rgb}{0.0, 0.8, 0.6}
\definecolor{cyan}{rgb}{0.0, 1.0, 1.0}
\definecolor{amber(sae/ece)}{rgb}{1.0, 0.49, 0.0}
\graphicspath{ {images/} }
\definecolor{vdarkred}{rgb}{0.6,0,0.2}

\usepackage[utf8]{inputenc}
\usepackage{lmodern}

\definecolor{vdarkred}{rgb}{0.6,0,0.2}
\definecolor{vdarkblue}{rgb}{0,0.2,0.6}
\usepackage[pdftex, colorlinks, linkcolor=vdarkblue,citecolor=vdarkred]{hyperref}

\author{	
J\"urg Fr\"ohlich
}

\date{August 2025}

\title{Chern-Simons theory in mathematics, condensed matter theory and cosmology\footnote{This review is based  
on the work of a variety of people who profited from mathematical insights and the generosity of Jim Simons, including  work 
by this author and his collaborators. It does not contain original results; but I hope it will still be somewhat useful and entertaining. 
I do not not provide an exhaustive list of references to original papers on topics mentioned; however, these papers can be easily 
found by tracing the literature quoted in the bibliography. I apologize to colleagues whose work should be quoted in this paper 
but is not.}}

\begin{document}

\maketitle

\begin{center}
{\large Dedicated to the memory of Jim Simons}
\end{center}

\vspace{1em}

\begin{abstract}
Various applications of Chern-Simons theory in algebraic topology, in particular knot theory, condensed matter physics 
and cosmology are reviewed. Special attention is paid to appearances of Chern-Simons actions in the theory of the 
(integer and fractional) quantum Hall effect. A mechanism related to five-dimensional abelian Chern-Simons theory 
that may be at the origin of the observed intergalactic magnetic fields in the universe is described.
\end{abstract}

\section{The Chern-Simons forms in mathematics}\label{Intro}

It is widely known and appreciated that the Chern-Simons forms and the corresponding Chern-Simons actions
play an interesting and important role in algebraic topology. Let $G$ be a compact Lie group.
Consider a principal $G$-bundle with connection $\nabla$ whose base space is a $(2n+2)$-dimensional manifold, $M$.
Let $A$ be the connection 1-form (gauge potential) with values in the Lie algebra of $G$, and let $F$ denote its curvature.
Let $\text{Tr}$ denote a trace on the Lie algebra of $G$ that is invariant under the adjoint action of $G$. Locally, the 
\textit{Chern-Simons $(2n+1)$-form,} $\omega_{2n+1}$, on $M$ is defined by the equation
$$d\omega_{2n+1}(A) = \text{Tr}\big(F^{\wedge (n+1)}\big)\,.$$
If $M$ is a compact manifold obtained by gluing together two oriented manifolds, $M_{+}$ and $M_{-}$, along their common
boundary, $\Lambda =\partial M_{+} = \partial M_{-}^{op}$, the \textit{Chern-Simons action} on $\Lambda$ is defined by
\begin{equation}\label{CS}
CS_{2n+1}^{\pm}(A) := \frac{1}{z_n \pi^{n}}\int_{M_{\pm}} \text{Tr}\big(F^{\wedge (n+1)}\big) = ``\frac{1}{z_n \pi^{n}}
\int_{\Lambda} \omega_{2n+1}(A)"\,,
\end{equation}
where $z_n$ is a certain integer ($z_1 = 4, z_2 =24, \dots$), and the right side is only defined modulo integers. 
One observes that
$$CS_{2n+1}^{+}(A)-CS_{2n+1}^{-}(A) = I\in \mathbb{Z}\,,$$
where $I$ is the ``instanton number'' associated with the connection $\nabla$ on the principal G-bundle over $M$.
It follows that the \textit{``Chern-Simons functional,''} $\text{exp}\big[2\pi\,i\,k\,CS_{2n+1}(A)\big], k\in \mathbb{Z},$ is a gauge-invariant, 
single-valued functional of the connection 1-form $A\big|_{\Lambda}$; $k$ is called its \textit{``level''}. Choosing an auxiliary Riemannian 
metric, $g$, on $M$, one may attempt (at least for $n=1$) to make mathematical sense of the functional integral
\begin{equation}\label{FI}
Z_{\Lambda}(k):=\int_{\mathcal{A}} \text{exp}\big[2\pi\,i\,k\,CS_{2n+1}(A)\big] \,\mathcal{D}_{g}A\,,
\end{equation}
where $\mathcal{A}$ is the affine space of gauge fields, and $\mathcal{D}_{g}A$ is a formal volume form on $\mathcal{A}$ 
that depends on the choice of the metric $g$. The dependence of the formal expression \eqref{FI} on $g$ is universal
and can be eliminated by dividing this functional by a standard one (e.g., a gravitational Chern-Simons partition 
function). For $n=1$, the functional integral in \eqref{FI} can be given a fairly precise mathematical meaning and shown 
to define a topological invariant of the framed 3-manifold $\Lambda$. Three-dimensional Chern-Simons theory turns out to
have interesting applications in the theory of knots and links embedded in $\Lambda$. Let $\chi_R$ be the character of
an irreducible representation, $R$, of the gauge group $G$, and let $\mathcal{K}$ be a framed knot embedded in $\Lambda$. A 
\textit{Wilson loop} ``observable,'' $W_R(\mathcal{K})$, associated with $\mathcal{K}$ is defined by
\begin{equation}\label{Wilson}
W_R(\mathcal{K};A):= \chi_R\Big(\mathcal{P}\text{exp}\Big[i \oint_{\mathcal{K}} A\Big]\Big)\,,
\end{equation}
where $\mathcal{P}$ denotes path ordering. One attempts to make sense of the expression
\begin{equation}\label{knots}
\Big<W_R(\mathcal{K})\Big>_k := Z_{\Lambda}(k)^{-1} \int_{\mathcal{A}} W_R(\mathcal{K};A)\cdot\text{exp}\big[2\pi\,i\,k\,CS_{3}(A)\big] \,
\mathcal{D}_{g}A\,,
\end{equation}
which formally defines an invariant of the framed knot $\mathcal{K}$ embedded in $\Lambda$.

There are various approaches to providing a precise mathematical meaning to this expression; see \cite{Witten, FK, Konts}
(the best known one being the one in \cite{Witten}). The approach described in \cite{FK} employs the following ingredients: 
Choosing $\Lambda = \mathbb{R}^{3}$, one may define the integrals in \eqref{FI} and \eqref{knots}
by fixing a special gauge, called \textit{``light-cone gauge.''} Let $x=(x^1, x^2, \tau)$ be Cartesian coordinates
on $\mathbb{R}^{3}$. We set $x^{+}=z:=x^1 +ix^2$, $x^{-}=\overline{z}:= x^1 -ix^2$, and write the gauge potential as
$$A(x)=a_{+}(x)\,dz\,+\,a_{-}(x)\,d\overline{z} + a_0(x)\,d\tau\,.$$
One chooses the gauge condition 
$$a_{-}(x)=0\,.$$
The Chern-Simons action then becomes a quadratic functional of $a_{+}$ and $a_0$, and the functional integrals in \eqref{FI} 
and \eqref{knots} become formal Gaussian integrals. The Wilson loop expectation $\Big<W_R(\mathcal{K})\Big>_k$ can be 
calculated by integration of \textit{Knizhnik-Zamolodchikov equations} \cite{KZ} (which are systems of ordinary
differential equations in the variable $\tau$), or by expanding the path-ordered exponential defining $W_R(\mathcal{K};A)$ 
in a power series and evaluating the formal Gaussian integrals. These are the methods employed in \cite{FK} 
to make sense of \eqref{knots}; they yield mathematically precise expressions for $\Big<W_R(\mathcal{K})\Big>_k$. 
For a somewhat more detailed review of these ideas see \cite{Fr-1}.

I will not present further details of applications of Chern-Simons theory to problems in algebraic topology. 
Instead, I propose to outline applications of three- and five-dimensional \textit{abelian} Chern-Simons theories ($G=\mathbb{R}$) 
to problems in condensed-matter physics and cosmology related to the quantum Hall effect and to physical systems in $3+1$ dimensions. 
In these applications, the Chern-Simons 3-form and 5-form are integrated over three- and five-dimensional manifolds, $\Lambda$,
respectively, with non-empty boundaries, $\partial \Lambda \not= \emptyset$. The resulting actions are \textit{anomalous}, i.e., 
they are not gauge-invariant. Their gauge variation is given by certain boundary terms, which are cancelled by the gauge variation of 
effective actions of chiral degrees of freedom attached to the boundary $\partial \Lambda$ that are coupled to the gauge field
$A\big|_{\partial \Lambda}$. These are insights going back to the analysis of anomalies in quantum gauge theories. 
They have interesting consequencs for chiral conformal field theory.

\section{Three-dimensional Chern-Simons theory and the Quantum Hall Effect} \label{QHE}
Let $\Lambda$ be an open domain in $\mathbb{R}^{3}$ with a non-empty boundary $\partial \Lambda$; (later,
$\Lambda$ will be an infinitely extended full cylinder whose axis is parallel to the time axis). Let $A$ be the electromagnetic
vector potential restricted to $\Lambda$. We consider the abelian Chern-Simons action
\begin{equation}\label{CS}
CS_{\Lambda}(A):=\frac{\sigma}{2} \int_{\Lambda} A \wedge F\,, \quad \text{ with }\,\, F=dA\,,
\end{equation}
where $\sigma$ is a real constant, which, later, will turn out to be the \textit{Hall conductivity.}

\textit{Remark:} If $\Lambda = \mathbb{R}^{3}$ and $F$ falls off rapidly at infinity then $CS_{\mathbb{R}^{3}}(A)$ is 
the so-called \textit{helicity} of the magnetic field, $\vec{B}$, dual to $F$. Helical magnetic fields (and helical fluid flows, 
with $A$ the velocity vector field of a fluid and $F$ dual to its vorticity) have been studied for a long time.

Consider a gauge transformation, $A\mapsto A+ d\chi$, where $\chi$ is a smooth function on $\mathbb{R}^{3}$;
one finds that
\begin{equation}\label{gauge-var}
CS_{\Lambda}(A+d\chi)= CS_{\Lambda}(A) + \frac{\sigma}{2} \int_{\partial \Lambda} \chi\, F\,,
\end{equation}
i.e., the Chern-Simons action on a manifold with boundary fails to be gauge-invariant; it is said to be \textit{``anomalous.''}

Let $a:=A\big|_{\partial \Lambda}$ be the restriction of the vector potential $A$ to the boundary, $\partial \Lambda$, of
a full cylinder $\Lambda \subset \mathbb{R}^{3}$ with axis parallel to the time axis, which we equip with a Lorentzian metric; and let $F=E_{\Vert}$ 
be the dual of $da$, i.e, the dual of the restriction of the field strength of the electromagnetic vector potential to the boundary 
of $\Lambda$. We consider the functional
\begin{equation}\label{chiral action}
\Gamma_{\partial \Lambda}(a) := \frac{1}{2} \int_{\partial \Lambda}d\text{vol}\,\big[\big(F- \text{div }a \big)
\Box^{-1}\big(F- \text{div }a\big) + a_{\mu}\,a^{\mu}\big]
\end{equation}
where $\Box$ is the d'Alembertian on $\partial \Lambda$. This is the anomalous effective action of a chiral current 
localized on $\partial \Lambda$ coupled to the electromagnetic vector potential $a$. The reader is invited to verify that the combination
\begin{equation}\label{anomaly cancel}
CS_{\Lambda}(A) - \frac{\sigma}{2} \Gamma_{\partial \Lambda}(a=A\big|_{\partial \Lambda})\,\, \text{ is gauge-invariant. }
\end{equation}
This observation implies that the gauge anomaly of the Chern-Simons action described in \eqref{gauge-var} 
is cancelled by the gauge anomaly of certain charged chiral degrees of freedom attached to the boundary of 
$\Lambda$ when they are coupled to the electromagnetic vector potential. (This will be made more precise below.)

Next, we propose to sketch how these functionals come up in the theory of the \textit{quantum Hall effect} (QHE). We consider 
a system consisting of a two-dimensional (2D) electron gas forming at the planar interface between a semi-conductor and 
an insulator when a gate voltage perpendicular to the interface is turned on pushing electrons from the bulk of the semi-conductor
to the interface. The ``space-time'' of such a gas is given by the full cylinder $\Lambda= \Omega \times \mathbb{R}$, where
$\Omega$ is the (bounded open) planar region the gas is confined to, and $\mathbb{R}$ is the time axis. Let $h$ denote 
Planck's constant, and $e$ the elementary electric charge; the density of the 2D electron gas is denoted by $n$. We imagine 
that the gas is put into a uniform magnetic field, $\vec{B}_0$, of strength $B_0$, measured in units of the flux quantum, $h/e$,
perpendicular to the plane of the electron gas. The dimensionless quantity $\nu:= \frac{n}{B_0}$ is called the 
``filling factor'' of the gas; it is equal to the number of Landau levels filled by a \textit{non-interacting} gas of 
electrons freely moving in $\Omega$ of density $n$.

We propose to consider the $(2+1)$-dimensional electrodynamics of this system. We assume that the filling 
factor $\nu$ is such that the ground-state energy of the gas is separated from the energy spectrum corresponding 
to conducting states by a strictly positive so-called mobility gap, so that the \textit{longitudinal (Ohmic) conductivity} of the gas 
\textit{vanishes} (at zero temperature).\footnote{To show that, for a 2D gas of \textit{interacting} electrons, there are filling factors,
$\nu$, where a mobility gap opens, is a difficult problem of quantum-mechanical many-body theory that is not particularly well 
understood, yet.} A 2D electron gas with these properties is called \textit{``incompressible.''} We study the response of 
an incompressible 2D electron gas to turning on a tiny, smooth, perturbing external electromagnetic field, 
$F$, where the coefficients, $\big\{F_{\mu\nu}\,\big|\, \mu, \nu =0,1,2\big\}$, of the 2-form $F$ are given by
\begin{equation}\label{field strength}
\big(F_{\mu\nu}\big) = \begin{pmatrix} 0&E_1&E_2\\-E_1&0&-B\\-E_2&B&0\end{pmatrix}\,.
\end{equation}
Here $\underline{E}=(E_1, E_2)$ is the component of the external electric field parallel to the plane of the 2D 
gas, and $B=B_{tot}-B_0$, where $B_{tot}$ is the component of the total external magnetic induction perpendicular 
to the plane of the system. (Note that the orbital motion of electrons confined to $\Omega$ only depends on
the components $\underline{E}$ and $B_{tot}$.)

The basic equations of the electrodynamics of a 2D incompressible electron gas (in units where the velocity of light 
$c=1$) are as follows.
\begin{enumerate}
\item[(i)] {\textbf{Hall's Law} -- \textit{a phenomenological law}
\begin{equation}\label{Hall law}
j^{k}(x)= \sigma_{H} \varepsilon^{k\ell}E_{\ell}(x)\,, \,\,\, k, \ell = 1,2,
\end{equation}
where  $\underline{j}=(j^1, j^2)$ denotes the electric current density of the gas, $\varepsilon^{12}=-\varepsilon^{21}= 1, \varepsilon^{ii}=0$, 
and $\sigma_H$ is the so-called \textit{Hall conductivity.}\footnote{Einstein's summation convention is used here and everywhere else in this text.}
Since the electron gas is assumed to be incompressible, an Ohmic contribution to the current density does \textit{not} appear 
on the right side of this equation; the \textit{longitudinal conductivity} of the gas \textit{vanishes}. 

Hall's law shows that if $\sigma_H \not= 0$ the symmetries of reflections in lines ($P$) and time reversal ($T$) are broken. 
These symmetries are broken explicitly by the presence of the external magnetic induction.}
\item[(ii)]{\textbf{Charge conservation} -- \textit{a fundamental law}
\begin{equation}\label{continuity eq}
\frac{\partial}{\partial t} \rho(x) + \underline{\nabla} \cdot \underline{j}(x) =0 \,,
\end{equation}
where $\rho$ is the electric charge density of the gas.
 }
 \item[(iii)]{\textbf{Faraday's induction law} -- \textit{a fundamental law}
 \begin{equation}\label{Faraday}
 \frac{\partial}{\partial t} B(x) + \underline{\nabla} \wedge \underline{E}(x)=0 \,,
  \end{equation}
and we have assumed that $B_0$ is constant.

Combining Eqs.~(i) through (iii), we obtain
  \begin{equation}\label{Rhodot}
  \frac{\partial}{\partial t}\rho \overset{(ii)}{=} -\underline{\nabla}\cdot \underline{j} \overset{(i)}{=} -\sigma_{H} \underline{\nabla} \wedge \underline{E} \overset{(iii)}{=} \sigma_{H} \frac{\partial}{\partial t} B\,.
 \end{equation}}
\item[(iv)]{\textbf{Chern-Simons Gauss law}\\
Integrating Eq.~\eqref{Rhodot} in time $t$, with integration constants chosen as follows
 $$j^{0}(x):= \rho(x) + e\cdot n, \quad B(x)= B_{tot}(x)- B_{0}\,,$$
with $B_{tot}$ the component of the (total) external magnetic induction and $B_0$ the component of the uniform 
magnetic induction in the direction perpendicular to $\Omega$, we find the so-called \textit{``Chern-Simons Gauss law''} 
 \begin{equation}\label{CS-Gauss}
 j^{0}(x)=\sigma_{H} B(x)\,.
 \end{equation}
This equation must be understood as saying that a small change, $\Delta B$, of the external magnetic induction 
perpendicular to the plane of the gas induces a small change, $\Delta j^{0}$, in the electric charge density of the gas.
Eq.\eqref{CS-Gauss} is also known under the name of \textit{Streda formula}; see \cite{Streda}.
Integrating both sides of Eq. \eqref{CS-Gauss} in time from $t=t_i$ to $t=t_f$ and over $\Omega$, we conclude that
\begin{equation}\label{charge transport}
\Delta Q= \sigma_H \Delta \Phi,
\end{equation}
where $\Delta Q$ is the change of the electric charge stored in the system, and $\Delta \Phi$ 
is the change of the magnetic flux through $\Omega$ during the time interval $[t_i,t_f]$.
}
\end{enumerate}
\vspace{0.1cm}

\textbf{Quantization of the Hall conductivity}: In order to explain the surprizing \textit{quantization} of the Hall conductivity $\sigma_H$, 
we consider an incompressible electron gas confined to a 2D torus, $\mathbb{T}^{2}=:\mathbb{T}$, and imagine threading  
a magnetic flux tube through the interior of $\mathbb{T}$ that does not cross/intersect $\mathbb{T}$ and is 
parallel to a non-contractible cycle, $\gamma$, of $\mathbb{T}$. One may ask how much electric charge, 
$\Delta Q_{\widetilde{\gamma}}$, crosses a cycle, $\widetilde{\gamma}$, on $\mathbb{T}$ conjugate to $\gamma$ 
(i.e., $\gamma$ and $\widetilde{\gamma}$ intersecting each other in a single point of $\mathbb{T}$) when the 
magnetic flux in the interior of the torus in the direction of $\gamma$ changes by an amount $\Delta \Phi$. Combining 
Faraday's induction law with Hall's law \eqref{Hall law}, we readily find that the quotient $\Delta Q_{\widetilde{\gamma}}/\Delta \Phi$ 
is given by the Hall conductivity $\sigma_H$. Assuming that all quasi-particles contributing to the electric current 
$\underline{j}$ in the system are \textit{electrons} or \textit{holes,} i.e., particles with electric charge $\mp e$, 
one can invoke the theory of the \textit{Aharonov-Bohm effect} to conclude that the state of this system is 
\textit{unchanged} if the magnetic flux in the interior of $\mathbb{T}$ is slowly increased by an \textit{integer multiple} of the 
flux quantum $h/e$. This implies that, in the process of increasing $\Delta \Phi$ by an amount $h/e$, an integer number, $N$, 
of electrons with a total electric charge -$Ne, N\in \mathbb{Z},$ must cross the cycle $\widetilde{\gamma}$. Hence
$$ \Delta Q_{\widetilde{\gamma}} = N e =\sigma_H (h/e) \quad \Leftrightarrow \quad \sigma_H = 
\frac{e^{2}}{h}\,N\,, \,\,N\in \mathbb{Z}\,,$$
which shows that, in an incompressible 2D electron gas \textit{without} fractionally charged quasi-particles contributing
to the electric current of the system, the Hall conductivity $\sigma_H$ must be an integer multiple of $\frac{e^2}{h}$. 
This reasoning process does \textit{not} explain why the \textit{``Hall fraction,''} $\frac{h}{e^{2}}\sigma_H$, is often 
observed \textit{not} to be an integer, but a \textit{rational number} (most often, but not always, with an odd denominator). 
It suggests, however, that if it is \textit{not} an integer then there must exist \textit{fractionally charged quasi-particles} 
contributing to the electric current in the system, as suggested by \textit{Tsui} and \textit{Laughlin} 
\cite{Tsui-Laughlin}.

The sign of $\sigma_H$ is determined by the sign of the electric charge of the quasi-particles carrying the Hall current, 
namely whether these quasi-particles are electrons or holes. Historically, this fact has led to the discovery of holes in 
nearly full conduction bands of semi-conductors.\\
\vspace{0.1cm}

\textbf{Quantum Hall effect and Chern-Simons action}: Let A be the vector potential on $\Lambda = \Omega \times \mathbb{R}$
with field strength (curvature) $F=dA$. Hall's law \eqref{Hall law} and the Chern-Simons Gauss law \eqref{CS-Gauss} 
yield the equation
\begin{equation}\label{general Hall}
\frac{\delta S_{eff}(A)}{\delta A_{\mu}} \equiv  j^{\mu}(x) = \sigma_{H} \varepsilon^{\mu\nu\lambda}F_{\nu\lambda}(x)\,,
\end{equation}
where $S_{eff}(A)$ is the so-called \textit{effective action} of the 2D electron gas; (the first equation in \eqref{general Hall} 
follows from the definition of effective actions). Using that $F=dA$, we can integrate \eqref{general Hall} to
conclude that, up to an irrelevant constant,
\begin{equation}\label{eff action}
S_{eff}(A) = \frac{\sigma_H}{2} \int_{\Lambda} A\wedge F\,,
\end{equation}
which is the \textit{Chern-Simons action} associated with the electromagnetic vector potential $A$. Eq.~\eqref{general Hall} 
is a \textit{generally covariant} relation between the current density, $j^{\mu}$, of the electron gas and the electromagnetic
field tensor $F_{\nu \lambda}$.

\textit{Remark:} It should be noted that the Chern-Simons action is \textit{not} the exact effective action of an incompressible
2D electron gas. But it is the leading term in an expansion of its effective action in a series of terms that are increasingly 
``irrelevant'' at large distance scales and low energies.\footnote{The term ``irrelevant'' is understood in the sense of standard
dimensional analysis.} The problem of predicting the form of 
effective actions of incompressible (``gapped'') electron gases  in two and three space dimensions has been dealt with in \cite{FST}. 

Eq.~\eqref{general Hall} seems to lead to a \textbf{contradiction}.
 \begin{equation}\label{puzzle}
 0\overset{(ii)}{=} \partial_{\mu}j^{\mu} \overset{(iii), \eqref{general Hall}}{=} \varepsilon^{\mu\nu\lambda}(\partial_{\mu} \sigma_{H}) F_{\nu\lambda}\,.
\end{equation}
We observe that, while the left side of this equation vanishes by charge conservation, the right side does 
\textit{not} vanish in general wherever the value of $\sigma_{H}$ jumps, as, for example, at the boundary, 
$\partial \Omega$, of the sample region $\Omega$ containing the 2D electron gas.

\textbf{Resolution of contradiction}: In Eq.~\eqref{general Hall}, $j^{\mu}$ is the \textit{bulk} current density, $j^{\mu}_{bulk}$, 
which is apparently \textbf{not} conserved, because it is \textit{not} the total current density! The \textit{conserved 
total electric current density,} $j^{\mu}_{tot}$, can be decomposed as
 \begin{equation}\label{edge currents}
 j^{\mu}_{tot}= j^{\mu}_{bulk} + j^{\mu}_{edge}, 
 \end{equation}
where $j^{\mu}_{edge}$ is an \textit{edge current density} whose support is equal to the support of $\underline{\nabla} \sigma_H$.
We then have that
 $$\partial_{\mu}j^{\mu}_{tot}=0, \,\,\,\text{   but   }\,\,\, \partial_{\mu}j^{\mu}_{bulk} =-\partial_{\mu}j^{\mu}_{edge} \overset{\eqref{puzzle}}{\not=} 0\,,$$
where
 $$\text{supp }j^{\mu}_{edge} = \text{supp}\lbrace \underline{\nabla} \sigma_{H} \rbrace \supseteq \partial \Omega, \qquad 
\text{with }\quad\,\, \underline{j}_{edge} \perp \underline{\nabla} \sigma_{H}\,.$$
Equations \eqref{puzzle} (setting $j^{\mu} = j^{\mu}_{bulk}$!) and \eqref{edge currents} imply that
\begin{equation*}
\partial_{\mu} j^{\mu}_{edge} \overset{\eqref{edge currents}}{=} -\partial_{\mu} j^{\mu}_{bulk}\vert_{\text{supp}\lbrace \underline{\nabla} \sigma_{H}\rbrace} \overset {\eqref{puzzle}}{=} - \varepsilon^{jk} \big(\partial_{j}\sigma_{H}\big) E_{k}\,.
\end{equation*}
When restricted to the support of $\underline{\nabla}\sigma_H$ this equation tells us that
\begin{equation}\label{2D anomaly}
\partial_{\mu} j^{\mu}_{edge} = -(\Delta \sigma_H) E_{\Vert}\big|_{\text{supp}\{\underline{\nabla}\sigma_H\}}\,,
\end{equation}
where $\Delta \sigma_H$ is the jump of the Hall conductivity across the edge in question, and $E_{\Vert}$ is the component
of the electric field parallel to the edge. This equation is an expression of the \textit{chiral anomaly} in 1+1 dimensions \cite{CAA}.
In the following, we assume that $\sigma_H$ is constant throughout the interior of 
the spatial domain $\Omega$ and vanishes outside $\Omega$ (so that $\Delta \sigma_H = \sigma_H$ in \eqref{2D anomaly}). 
This is not a realistic assumption; but it simplifies our discussion without introducing misleading concepts.
Equation \eqref{2D anomaly} shows that $j^{\mu}_{edge}$ is an \textit{anomalous chiral} current density in $1+1$ dimensions. 
It can be viewed as a manifestation of \textit{``holography''}: The degrees of freedom propagating along the edge of the 2D electron
gas store as much information about the Hall conductivity and other quantities characteristic of the system as the degrees
of freedom in the bulk. (A prediction of chiral edge currents in 2D electron gases exhibiting the quantum Hall effect 
first appeared in \cite{Halperin}.)

Not surprisingly, the observations made here are closely related to ones that we have made before, namely in Eqs.~\eqref{gauge-var}, 
\eqref{chiral action} and \eqref{anomaly cancel}: The Chern-Simons action \eqref{CS}, with $\sigma= \sigma_H$ 
the Hall conductivity, on a three-dimensional space-time manifold $\Lambda = \Omega\times \mathbb{R}$ with a non-empty 
boundary, $\partial \Omega \times \mathbb{R}$, is not gauge-invariant. Its gauge anomaly is cancelled by the one of the 
action functional $-\frac{\sigma_H}{2} \Gamma_{\partial \Lambda}(a)$ introduced in \eqref{chiral action}, which is the effective action 
of chiral charged degrees of freedom propagating along the boundary $\partial \Omega$ of the sample domain $\Omega$
of the 2D gas, which give rise to the chiral edge current $j^{\mu}_{edge}$. (Related observations have been made in \cite{Wen}.)

The theory of chiral degrees of freedom in 1+1 dimensions leads to predictions of the possible values of the Hall conductivity $\sigma_H$
and of the spectrum of quasi-particles exhibited by the electron gas. A general approach to this topic (based on the
theory of super-selection sectors of a class of chiral conformal field theories in 1+1 dimensions) has been presented in \cite{FPSW}. 
It predicts that $\sigma_H$ is a \textit{rational} multiple of $\frac{e^2}{h}$ and that a 2D electron gas with a Hall conductivity 
that is \textit{not} an integer multiple of $\frac{e^2}{h}$ exhibits quasi-particles with fractional electric charge and exotic 
quantum statistics (called \textit{fractional} or \textit{braid} (goup) \textit{statistics}, \cite{Fr-2}).\footnote{Such quasi-particles, 
which, in the bulk, are quasi-static, might be of interest for topological quantum computing.} Here I only mention a special case of 
the general theory treated in \cite{Fr-1} and \cite{FST} (and references given there), which is relevant for the interpretation of a very 
large set of experimental data. 
I suppose that the chiral electric edge current density, $j^{\mu}_{edge}$, is a linear combination of several (canonically 
normalized) chiral current densities, $j^{\mu}_{\alpha}, \,\alpha=1,\dots, N$, associated with ``emergent'' Kac-Moody symmetries
(at level 1) acting on the space of edge states,
\begin{equation}\label{egde channels}
j^{\mu}_{edge}= \sum_{\alpha=1}^{N} Q^{\alpha}\,j^{\mu}_{\alpha}\,,
\end{equation}
where $Q^{1}, \dots, Q^{N}$ are real numbers. The current density $Q^{\alpha}\,j^{\mu}_{\alpha}$ is the contribution of chiral
degrees of freedom in an ``edge channel,'' labelled $\alpha$, to the electric edge current density $j^{\mu}_{edge}$, for 
$\alpha=1, \dots, N$. (For an incompressible 2D gas of \textit{non-interacting} electrons, these channels correspond 
to the edge states of $N$ filled Landau levels, with $Q_{\alpha}=1,$ for $\alpha=1,\dots, N$.) The theory of the chiral 
anomaly in two dimensions \cite{CAA} tells us that
\begin{equation}\label{Hall cond}
\partial_{\mu} j^{\mu}_{edge} = \sum_{\alpha=1}^{N} Q^{\alpha}\partial_{\mu}j^{\mu}_{\alpha} = - \frac{e^{2}}{h} \big[ \sum_{\alpha} (Q^{\alpha})^{2}\big] E_{\Vert}\,,
\end{equation}
where $E_{\Vert}$ is the component of the external electric field parallel to $\partial \Omega$. 
Together with \eqref{2D anomaly} (setting $\Delta \sigma_H = \sigma_H$), this equation yields a formula for $\sigma_H$, 
namely
\begin{equation}\label{Hall cond}
 \sigma_{H}= \frac{e^{2}}{h} \sum_{\alpha} (Q^{\alpha})^{2}\,.
 \end{equation}
One can show (see \cite{Fr-1}, \cite{FST} and references given there) that the quantum numbers of chiral edge states describing 
configurations of electrons and holes propagating along the edge, $\partial \Omega$, of the 2D electron gas clockwise or 
anti-clockwise (depending on the sign of $B_0$) form an $N$-dimensional \textit{odd integral lattice,} $\mathcal{L}$, 
and that $\underline{Q}:=\big(Q^1, \dots, Q^{N}\big)$ is a \textit{visible vector} in the dual lattice $\mathcal{L}^{*}$. 
This implies that $\sum_{\alpha} (Q^{\alpha})^{2}$ is a \textbf{rational number}, i.e., $\sigma_H$ is a \textbf{rational multiple} 
of $\frac{e^{2}}{h}$. The sites in the lattice $\mathcal{L}^{*}$ are the quantum numbers of configurations of quasi-particles 
some of which are fractionally charged and have fractional statistics; (except when $\mathcal{L}=\mathcal{L}^{*}$, which holds,
e.g., for non-interacting 2D electron gases).

The theory sketched here can also be developed by focusing on the physics in the bulk of the 2D electron gas. In 2+1 dimensions, 
a conserved current density, $j$, is dual to a closed 2-form, $J$. If the space-time, $\Lambda$, of the gas is a full cylinder then $J$ 
is exact, i.e., $J=dZ$ (Poincar\'e's lemma), where $Z$ is a 1-form, the \textit{``vector potential''} of the conserved current density. 
In a 2D incompressible electron gas (which is a so-called ``topological insulator''), the \textit{quantum theory} of a conserved current 
at very large distance scales and low energies can be described by functional integrals with a ``topological'' action functional, 
$S_{\Lambda}(Z)$, depending on the vector potential $Z$, given by 
$$S_{\Lambda}(Z)=\frac{h}{2}\Big\{\int_{\Lambda} Z\wedge dZ - \Gamma_{\partial \Lambda}(z=Z\big|_{\partial \Lambda})\Big\},$$
i.e., by a Chern-Simons action. Here $\Gamma_{\partial \Lambda}$ is a boundary term restoring invariance under 
``gauge transformations,'' $Z \mapsto Z+ d\zeta$, with $\zeta$ a real-valued function on $\Lambda$; see \eqref{chiral action}. 
For an incompressible gas with $N$ conserved current densities, $J_{\alpha} = dZ_{\alpha}, \,\, \alpha=1,\dots, N\,,$
the action functional is given by
$$S_{\Lambda}(\underline{Z})=\frac{h}{2}\sum_{\alpha=1}^{N} \Big\{\int_{\Lambda} Z_{\alpha}\wedge dZ_{\alpha} -
\Gamma_{\partial \Lambda}(Z_{\alpha}\big|_{\partial \Lambda})\Big\}\,, \quad \text{ with }\,\,\underline{Z}=\big(Z_1, \dots, Z _N\big)\,.$$
Coupling the electric current density $j^{\mu}=\sum_{\alpha=1}^{N}Q^{\alpha}j^{\mu}_{\alpha}$ to the vector potential, $A$, of an 
external electromagnetic field by adding the term 
$$e\,J\wedge A= e\sum_{\alpha=1}^{N} Q^{\alpha}\, dZ_{\alpha} \wedge A\,,$$ 
to the action, we find the \textit{effective action} of the system by carrying out the functional integral
\begin{equation}\label{Eff Action}
e^{iS_{eff}(A)/\hbar} :=\text{const.}\int \prod_{\alpha=1}^{N} \mathcal{D}Z_{\alpha}\,\text{exp}\Big[\frac{i}{\hbar}\Big(S_{\Lambda}(\underline{Z})
+ \int_{\Lambda} eJ \wedge A \Big)\Big]
\end{equation}
(with the constant chosen such that $S_{eff}(A=0)=0$) and find that
$$S_{eff}(A)= \frac{e^{2}}{2h}\sum_{\alpha=1}^{N} (Q^{\alpha})^{2}\Big\{ \int_{\Lambda} A\wedge F  
- \Gamma_{\partial \Lambda}\big(A\big|_{\partial \Lambda}\big)\Big\}\,.$$
This reproduces formula \eqref{eff action}, with $\sigma_H$ given by \eqref{Hall cond}, and it shows 
that one obtains the same expression for the Hall conductivity independently of 
whether one considers the physics of edge states or the bulk of an incompressible 2D electron gas (an instance of ``holography'').

\textit{Remark:} An interesting \textit{duality} between 2D \textit{insulators} and 2D \textit{superconductors} and a 
\textit{self-duality} of 2D incompressible electron gases exhibiting the quantum Hall effect is revealed by functional 
Fourier transformation, as in \eqref{Eff Action}, with the electromagnetic vector potential $A$ and the vector potential 
$Z$ of the conserved electric current density $J$ playing the role of conjugate variables \cite{FST}.

What we have outlined here is how the theory of \textit{``abelian'' quantum Hall fluids} is intimately related to \textit{abelian Chern-Simons 
theory} of the vector potentials of exact current 2-forms; (a more general theory, implicitly related to non-abelian Chern-Simons theory, 
is described in \cite{FPSW}). The facts that quantum numbers of configurations of quasi-particles correspond to sites in the dual, 
$\mathcal{L}^{*}$, of an odd integral lattice $\mathcal{L}$, and that $\underline{Q}=\big(Q_1, \dots, Q_N\big)$ is a ``visible'' vector 
in $\mathcal{L}^{*}$ lead to plenty of very specific predictions of properties of 2D incompressible electron gases exhibiting 
the quantum Hall effect that match \textit{experimental data} with astounding accuracy; (this theme is developed in work
reviewed in \cite{Fr-1}, \cite{FST}, and refs.).

\section{The five-dimensional Chern-Simons action and a cousin of the quantum Hall effect}\label{5D QHE}
This section closely follows ideas described in \cite{FP} and reviewed in \cite{Fr-3}. We consider a cousin of the quantum 
Hall effect in systems of charged matter on a five-dimensional (5D) space-time slab, $\Lambda=\Omega\times [0,L]$, 
with two four-dimensional ``boundary branes,'' $\partial_{\pm}\Lambda\simeq \Omega$, parallel to the 
$(x^0, x^1, x^2, x^3)$-plane in 5D Minkowski space, where $x^0$ is the time coordinate.
The slab $\Lambda$ is assumed to be filled with \textit{very heavy, four-component Dirac fermions} coupled to the 5D 
electromagnetic vector potential, $\widehat{A}$. Functional integration over configurations of fermion degrees of 
freedom yields an effective action depending on $\widehat{A}$ that has the form
\begin{align}\label{8.9}
\begin{split}
S_{\Lambda}(\widehat{A})=& \frac{1}{2 L\alpha} \int_{\Lambda} d^{5}x\,\widehat{F}_{MN}(x) \widehat{F}^{MN}(x) + 
CS_{\Lambda}(\widehat{A})\\
- & \Gamma_{\ell}(\widehat{A}\vert_{\partial_{+}\Lambda}) - \Gamma_{r}(\widehat{A}\vert_{\partial_{-}\Lambda})+\cdots \,,
\end{split}
\end{align}
where $M,N=0,\dots,4, \, \alpha$ is a dimensionless constant, $L$ is the width of the 5D slab $\Lambda$, and
\begin{equation}\label{8.10}
CS_{\Lambda}(\widehat{A}):= \frac{\kappa_H}{24 \pi^{2}} \int_{\Lambda} \widehat{A} \wedge \widehat{F} \wedge 
\widehat{F}
\end{equation}
is the \textit{5D Chern-Simons action}; (henceforth Planck's constant $\hbar$ is set to 1); the dots on the right side of
\eqref{8.9} stand for terms that are ``irrelevant'' on large distance scales and at low energies.
The vector potential $\widehat{A}$ is treated as a classical external field. The action \eqref{8.10} yields a formula for a
5D analogue of the Hall current (see \eqref{Hall law}):
\begin{equation}\label{8.10'}
j^{M}= \frac{\kappa_H}{8\pi^{2}}\, \varepsilon^{MNJKL}F_{NJ}\,F_{KL}\,,
\end{equation}
where $\kappa_H$ is a dimensionless constant, with $\kappa_H \in \mathbb{Z}$ (assuming  that all fermions 
have electric charges that are integer multiples of the elementary electric charge); it plays the role of the Hall fraction 
$\frac{h}{e^{2}}\sigma_H$. In the following $\kappa_H$ is set to 1. Equation \eqref{8.10'} is the five-dimensional analogue  
of equation \eqref{general Hall} in Sect.~2. 

The action functional in equation \eqref{8.9}, with $CS_{\Lambda}(\widehat{A})$ as in \eqref{8.10}, is the effective action 
describing a 5D analogue of the quantum Hall effect. Apparently, such effects can be observed in certain systems of 
condensed-matter physics with some ``virtual dimensions'' playing the role of space dimensions; see \cite{Zilber} and 
references given there.

In \eqref{8.9}, the functionals $\Gamma_{\ell/r}$ are the anomalous effective actions of left-handed/right-handed 
Dirac-Weyl fermions propagating along the boundary branes, $\partial_{\pm}\Lambda$.
The gauge anomaly of the actions $-\Gamma_{\ell}$ and $-\Gamma_r$ on the right side of \eqref{8.9} cancels the 
one of the Chern-Simons action $CS_{\Lambda}(\widehat{A})$; see \cite{CAA}.
 
It is of interest to study the \textit{dimensional reduction} of the 5D theory, with effective action given in
\eqref{8.9} and \eqref{8.10}, to a four-dimensional space-time. We consider gauge fields $\widehat{A}$ with the property
that, for an appropriate choice of gauge, the components $\widehat{A}_M$ are \textit{independent} of $x^{4}$, for $M=0,1,2,3$.
  A field $\varphi$ of scaling dimension 0, henceforth called \textit{axion} field, is defined by
$$\varphi(x):= \int_{\gamma_x}\widehat{A}\,,$$
where $\gamma_x$ is a straight curve parallel to the $x^{4}$-axis connecting $\partial_{-}\Lambda$ to $\partial_{+} \Lambda$, 
with $x:=(x^{0}, x^{1}, x^{2}, x^{3})$ kept fixed. The action functional in \eqref{8.9} then becomes
\begin{align}\label{8.11}
S_{\Omega}(A;\varphi)=&\frac{1}{2\alpha} \int_{\Omega} d^{4}x\big[ F_{\mu\nu}(x)F^{\mu\nu}(x) + \frac{1}{L^{2}}\partial_{\mu}\varphi(x)\partial^{\mu}\varphi(x) \big]\nonumber\\
+&\frac{1}{8\pi^{2}}\int_{\Omega} \varphi\,(F\wedge F) - \Gamma_{\Omega}(A)+\cdots\,, \quad \mu, \nu =0,\dots,3\,.
\end{align}
Under the present assumptions on the vector potential $\widehat{A}$ 
the boundary effective action $\Gamma_{\Omega}(A)=\Gamma_{\ell}(A)+\Gamma_{r}(A)$, with $A=\widehat{A}\big|_{\partial \Lambda}$, 
is \textit{not} anomalous and can be ignored in the following.

Expression \eqref{8.11} shows that the pseudo-scalar field $\varphi$ can indeed be interpreted as an 
\textit{axion} field; see \cite{axion}. One can add a self-interaction term, $U(\varphi)$, to the Lagrangian density in 
\eqref{8.11}, requiring that $U(\varphi)$ be periodic in $\varphi$ with period $2\pi$; such self-interaction terms can 
be induced by coupling the axion $\varphi$ to massive non-abelian gauge fields, which are subsequently integrated out. 
From \eqref{8.11} we derive the equations of motion for the electromagnetic field tensor $F_{\mu\nu}$ and the axion $\varphi$,
\begin{equation}\label{axion ED}
\partial_{\mu}F^{\mu\nu}= -\frac{\alpha}{4\pi^{2}} \partial_{\mu}\big(\varphi \widetilde{F}^{\mu\nu}\big),\qquad 
L^{-2}\,\Box\,\varphi=\frac{\alpha}{8\pi^{2}} F_{\mu\nu}\widetilde{F}^{\mu\nu}-\frac{\delta U(\varphi)}{\delta \varphi}\,, 
\end{equation}
where $\widetilde{F}^{\mu\nu}$ is the dual field tensor. These are the field equations of \textit{``axion-electrodynamics''} in four 
dimensions. In terms of electric and magnetic fields, $\vec{E}$ and $\vec{B}$, these equations are given by
\begin{align}\label{8.12}
\vec{\nabla}\cdot \vec{B}=&\,\,0\,, \quad \vec{\nabla}\wedge \vec{E} + \dot{\vec{B}}=0\,, \nonumber \\
\vec{\nabla}\cdot \vec{E}=&\,\, \frac{\alpha}{4\pi^{2}}\big(\vec{\nabla}\varphi\big)\cdot \vec{B}\,, \\
\vec{\nabla}\wedge \vec{B}=&\,\, \dot{\vec{E}} + \frac{\alpha}{4\pi^{2}}\lbrace \dot{\varphi}\vec{B} +
\vec{\nabla}\varphi\wedge \vec{E} \rbrace\,, \nonumber
\end{align}
and
\begin{equation}\label{8.12'}
L^{-2}\,\Box\,\varphi = \frac{\alpha}{4 \pi^{2}} \vec{E}\cdot \vec{B} - \frac{\delta U(\varphi)}{\delta \varphi} ,
\end{equation}
(we use units where the velocity of light is set to 1). Comparing the last equation in \eqref{8.12} with Maxwell's
generalization of Amp\`ere's law, we realize that 
\begin{equation}\label{current}
\vec{j}:=\frac{\alpha}{4\pi^{2}}\big\{ \dot{\varphi}\vec{B} +\vec{\nabla}\varphi\wedge \vec{E}\big\}
\end{equation}
can be interpreted as the electric current density of the system (ignoring a term corresponding to an Ohmic current).

We conclude this review by mentioning some interesting applications of axion-electrodynamics in condensed matter
physics and cosmology.

\subsection{Chiral magnetic effect and 3D quantum Hall effect}
We begin by considering special axion field configurations where $\varphi$ only depends on time $t=x^0$.
We set $\mu_5\equiv \mu_{\ell}- \mu_r := \dot{\varphi}$. In the original formulation of Eqs.~\eqref{8.9}, \eqref{8.10} and \eqref{8.10'}
the quantities $\mu_{\ell}$ and $\mu_r$ can be interpreted as the chemical potentials of the boundary branes $\partial_{+}\Lambda$
and $\partial_{-}\Lambda$, resp.; i.e., $\mu_5$ is the ``voltage drop'' between the two boundary branes.
The electric current density is then given by
\begin{equation}\label{top current}
\vec{j} = \frac{\alpha}{4\pi^{2}} \mu_5 \vec{B}\,,
\end{equation}
which is the equation describing the so-called \textit{chiral magnetic effect} \cite{CME}. 

In condensed-matter theory, the equation of motion of $\dot{\varphi}\equiv \mu_5$ takes the form of a 
\textit{diffusion equation},
\begin{equation}\label{8.14}
\dot{\mu}_{5} + \tau^{-1}\mu_{5} - D\bigtriangleup \mu_{5} = \frac{L^{2}\alpha}{4\pi^{2}} \vec{E} \cdot \vec{B}\,,
\end{equation}
where $\tau$ is a \textit{relaxation time} associated with processes mixing left- and right-handed quasi-particles in
certain solids, dubbed \textit{Weyl semi-metals,} $D$ is a \textit{diffusion constant}, and $\alpha = \frac{2\pi e^{2}}{h}$ is 
the fine structure constant; (it is assumed here that $U(\varphi)\equiv 0$). 
Eq.~\eqref{8.14} implies that $\mu_{5}$ approaches $\mu_{5}\simeq \tau\frac{L^2 \alpha}{4\pi^{2}} \vec{E}\cdot \vec{B}\,,$
as time $t$ tends to $\infty$ ($\vec{E}\cdot \vec{B}$ is assumed to be slowly varying in space, so that $\mu_5$ 
is nearly space-independent). 
This expression for $\mu_5$ can be plugged into equation \eqref{top current} for the vector current density $\vec{j}$ predicted by
the chiral magnetic effect. By comparing the resulting equation with \textit{Ohm's law} we find an expression for the 
\textit{conductivity tensor,} 
$\sigma=\big(\sigma_{k\ell}\big)$, namely
\begin{equation}\label{8.16}
\sigma_{k\ell}=\tau \big(\frac{L\alpha}{4\pi^{2}}\big)^{2} B_{k}B_{\ell}\,.
\end{equation}
This expression is relevant for studies of charge transport in Weyl semi-metals. 

Next, I consider a 3D \textit{spatially periodic} (crystalline) system of matter exhibiting (among others) magnetic
degrees of freedom that can be described by an ``emergent'' axion field, $\varphi$. The  crystal lattice is denoted by $\mathcal{L}$. 
The effective action, $S_{\Omega}\big(A;\varphi\big)$, of the system is given by \eqref{8.11}. Imposing periodic boundary conditions 
at the boundary, $\partial \Omega$, of space-time $\Omega$, the functional $\text{exp}\big[i S_{\Omega}\big(A;\varphi\big)\big]$ 
turns out to be periodic in $\varphi$ under shifts $\varphi \mapsto \varphi + 2n\pi, n \in \mathbb{Z}$. 
We consider a \textit{stationary state} of the system, with $\varphi$ time-independent, i.e., $\mu_5 =0$, and we assume 
that the state of the system is invariant under (lattice) translations leaving $\mathcal{L}$ invariant. Then the ``axion field''
$\varphi$ must be invariant under lattice translations, modulo $2\pi$; hence it is given by
\begin{equation}\label{9.7}
\varphi(\vec{x})=2\pi \,\big(\vec{K}\cdot \vec{x}\big) + \phi(\vec{x})\,,
\end{equation}
where $\vec{K}$ is a vector in the \textit{dual lattice,} $\mathcal{L}^{*}$, and $\phi$ is a function that is invariant under 
lattice translations, which we neglect. Using that $ \vec{\nabla}\varphi= 2\pi \vec{K}$, we find that Eqs.~\eqref{8.12} and \eqref{current}
imply the following formulae for the electric charge density, $\rho$, and current density, $\vec{j}$,
$$\rho= \frac{e^2}{h} \vec{K}\cdot \vec{B}\,, \qquad \vec{j}= \frac{e^{2}}{h}\vec{K}\times \vec{E}\,, \qquad \vec{K}\in \mathcal{L}^{*}\,.$$
This is Halperin's 3D quantum Hall effect \cite{Halperin-2}.

There are numerous further applications of the ideas sketched here in condensed matter physics; 
(see, e.g., \cite{Fr-3} for a review with plenty of references to orginal papers).

\subsection{The generation of primordial magnetic fields in the universe}
It is interesting to study models of the universe featuring an axion field $\varphi$, besides the electromagnetic field (and other
degrees of freedom describing visible matter). Ultralight axions may give rise to \textit{dark matter} (see, e.g., \cite{Peccei} 
and references given there), which produces observable gravitational effects, such as anomalies in rotation curves or 
gravitational lensing. In the following we study axion electrodynamics for the purpose of sketching a mechanism that might
give rise to the intergalactic magnetic fields observed in the universe. The equations given in \eqref{8.12} and \eqref{8.12'} 
must then be modified so as to account for the curvature of space-time. To keep our discussion simple, we suppose that 
the large-scale structure of the universe is described by a conformally flat \textit{Friedmann-Lema\^itre space-time,} $\Lambda$, 
with a metric, $(g_{\mu\nu})$, given by $g_{00}\equiv 1,\, g_{0j}\equiv 0,$ and\, $g_{ij} = -a(t)^{2} \delta_{ij},$ for $i,j = 1,2,3,$ 
where $t$ is cosmological time, and $a(t)$ is the scale factor. 
The Lorentzian geometry of $\Lambda$ is encoded in the \textit{Hubble ``constant,''} $H=\frac{\dot{a}}{a}$, a function 
of cosmological time expressing the observed expansion of the universe. 

For a Friedmann-Lema\^itre universe it is straightforward to find the modifications due to curvature 
of the equations of axion electrodynamics given in \eqref{8.12} and \eqref{8.12'}. The electric field $\vec{E}$ can be
eliminated by using the first three equations in \eqref{8.12} and plugging the result into the last equation. For simplicity
we assume that the axion field $\varphi$ is very slowly varying in space, so that terms proportional to 
$\vec{\nabla} \varphi$ can be neglected. Then the field equation for the magnetic induction $\vec{B}$ is given by
\begin{align}\label{magnetic induction}
\ddot{\vec{B}} -\Delta \vec{B} + 3H \dot{\vec{B}} + [\frac{3}{2}\dot{H} + (\frac{3}{2} H)^{2}] \vec{B} 
- \frac{\alpha}{4\pi^{2}}\, \dot{\varphi}\,\vec{\nabla}\wedge \vec{B} \, = \, 0.
\end{align}
Here and in the last equation of \eqref{8.12} we have neglected an Ohmic contribution to the vector current density. 
This is justified by noticing that, after recombination of charged particles into neutral atoms and molecules, Ohmic currents 
are tiny. We solve Eq.~\eqref{magnetic induction} by Fourier transformation in the spatial variables $\vec{x}=(x^1, x^2 , x^3)$,
\begin{equation}\label{B}
\vec{B}(\vec{x}, t)= \int_{\mathbb{R}^{3}}d^{3}k\,\,\vec{\mathcal{B}}(\vec{k})\,e^{i(\vec{k}\cdot \vec{x}-\omega(k)\, t)}\,,
\end{equation}
where $\vec{k}\cdot \vec{\mathcal{B}}(\vec{k})=0$, because $\vec{\nabla}\cdot \vec{B} =0$, and $k:=\vert \vec{k}\vert$. 
Without loss of generality, we may suppose that $\mu_5=\dot{\varphi}\gg H \geq 0$ during a certain interval $\mathcal{I}$ 
of cosmological time in the evolution of the universe; ($\mu_5<0$ can be treated similarly). To simplify matters, we consider 
an approximate solution of \eqref{magnetic induction} valid under the assumptions that $H$ and  $\mu_5$ are slowly 
varying functions of time $t\in \mathcal{I}$, for a range of $\vec{k}$-vectors indicated in \eqref{shell of growth} below. 
Plugging \eqref{B} into \eqref{magnetic induction}, we find that
\begin{equation}\label{frequency}
\omega(k)=-i\frac{3}{2}H  \pm \sqrt{ k^{2} + \frac{3\dot{H}}{2} \pm \mu_{5} k}\,.
\end{equation}
This formula shows that the expansion of the Universe ($H>0$) leads to power-law (actually, for a constant $H$, exponential)
damping of $\vec{B}$ in time if its Fourier amplitude, $\vec{\mathcal{B}}$, is supported outside of the shell in $\vec{k}$-space 
given by
\begin{equation}\label{shell of growth}
\Sigma := \Big\{\vec{k}\, \big|\, \mu_5-\sqrt{\mu_{5}^2-K}<  2k < \mu_5 + \sqrt{\mu_{5}^2-K}\Big\}\,,\quad \text{with }\,\, K:=9H^2 + 6\dot{H}\,.
\end{equation}
However, if the support of the Fourier amplitude $\vec{\mathcal{B}}$ of the magnetic induction intersects the shell $\Sigma$ 
specified in \eqref{shell of growth} then $\vec{B}$ grows \textit{exponentially} in time $t$. The magnetic induction develops a 
non-vanishing helicity, $\int \vec{A}\cdot \vec{B}$, where $\vec{A}$ is the electromagnetic vector potential.

Actually, $\mu_5 = \dot{\varphi}$ is of course \textit{not} constant in time, but dies out, $\mu_5 \rightarrow 0$, at large times $t$,
so that the exponential growth of $\vec{B}$ comes to an end. The conversion of axion oscillations, $\mu_5\not=0$, into
a helical magnetic induction with an ever growing wave length $\lambda$ (initially given by $\lambda = 2\pi\,|\vec{k}|^{-1},$
with $\vec{k} \in \Sigma$) could explain the observed presence of tiny, very homogeneous intergalactic magnetic fields 
in the universe. More details on these matters can be found in \cite{TW, JS, FP, Durrer}.\\

To conclude, I hope to have convinced readers that, besides the role they play in algebraic topology, Chern-Simons forms 
and -actions appear prominently in the theoretical interpretation of fascinating effects in condensed matter physics and 
cosmology.\\

\textbf{Acknowledgements}. I have greatly profited from numerous discussions with my colleagues and
friends, the late V.~F.~R.~Jones (braid groups and knot theory), L.~Michel (integral lattices) and R.~Morf 
(quantum Hall effect). I have had the great privilege of joint efforts with many people, including my friend Chr.~King, 
who passed away too soon, my former PhD students T.~Kerler, B.~Pedrini and U.~M.~Studer, my postdocs 
E.~Thiran and Chr.~Schweigert, and my colleague A.~Zee, besides further people. I thank J.-P.~Bourguignon for 
a very careful reading of the manuscript and for suggesting corrections.

I gladly recall that I was invited to present material closely related to what is described in Sections 1 and 2 in a lecture at Stony Brook 
on the occasion of Jim Simons' $60^{\text{th}}$ birthday, back in 1998. -- I am grateful to J.~Cheeger and B.~Lawson for 
giving me the opportunity to contribute this paper to the Memorial Collection in honor of Jim Simons.

\begin{center}
-----
\end{center}

\bigskip

\noindent
J\"urg Fr\"ohlich, ETH Z\"urich, Institute for Theoretical Physics, \href{mailto:juerg@phys.ethz.ch}{juerg@phys.ethz.ch}.
\\[0.3em]

\end{document}